\documentclass[10pt]{article}
\begin{document}

\title{Deformations of the geometry of  lipid vesicles}

\author{
R. Capovilla${}^{(1)}$, J. Guven${}^{(2)}$ and J.A. Santiago${}^{(2)}$ \vspace{1cm}\\
${}^{(1)}${\it Departamento de F\'{\i}sica}\\
{\it Centro de Investigaci\'on y de Estudios Avanzados del IPN}\\
{\it Apdo. Postal 14-740,07000 M\'exico, DF, MEXICO}
\vspace{.5cm}\\
${}^{(2)}${\it Instituto de Ciencias Nucleares}\\
 {\it Universidad Nacional Aut\'onoma de M\'exico}\\
{\it Apdo. Postal 70-543, 04510 M\'exico, DF, MEXICO}
}
\maketitle

\begin{abstract}
Consider a closed lipid membrane (vesicle), modeled as a two-dimensional
surface, described by a geometrical 
hamiltonian that depends on its extrinsic curvature.
The vanishing of its first variation determines the equilibrium configurations
for the system. In this paper, we examine 
the second variation of the hamiltonian about any given equilibrium, using an
explicitly surface covariant geometrical approach. We identify the operator 
which determines the stability of equilibrium configurations.
\end{abstract}


\section{Introduction}

Many of the physical properties of a lipid membrane (or vesicle) are captured
by a hamiltonian which describes the degrees of freedom of  
an idealized two-dimensional membrane surface. 
Typically, this hamiltonian is 
a sum of terms each of which either penalizes or constrains some
geometrical characteristic of this surface.
A term quadratic in the mean extrinsic
curvature is a measure of the energy penalty associated 
with the bending of the membrane \cite{Can:70,Hel:73,Eva:74}.
This term is supplemented by various constraints: the area
as well as the enclosed volume are usually fixed. A constraint or (a term in the energy) linear 
in the mean extrinsic curvature describes an asymmetry
between the lipid bilayers which constitute the membrane \cite{Sve.Zek:89}.
(For a review see
\cite{Nel.Pir.Wei:89,Saf:94,Pel:94,Sve.Zek:96,Sei:97}.)

In equilibrium, this hamiltonian will be stationary with respect to arbitrary
infinitesimal deformations in the surface. This 
infinitesimal deformation has both a tangential part and a normal part.
The former corresponds to a reparametrization of the surface, 
and so can contribute only in boundary terms. 
The remaining normal infinitesimal deformation will completely describe the 
physical state of deformation of the membrane in its bulk.
As a consequence, the equilibrium condition is a single
equation, known somewhat prosaically as `the shape equation' 
\cite{Hel.OuY:87}. 
In the thirty or so years since the introduction of the model,
a lot has been learnt about the solutions of this equation,
both analytically and numerically (see, for example,
\cite{Sei:97} and the references it contains).

In this paper, we will 
examine the second variation of the hamiltonian. 
From one point of view, the calculation of the second variation
about an equilibrium configuration is 
necessary in order to assess its degree of 
stability; in addition,
the fluctuations of the membrane 
due to its interaction with its environment, 
are encoded completely at lowest order
in the second variation of the hamiltonian.
This problem has, of course, been addressed before and in both contexts.
In the former, for example, perturbations about
spherical or cylindrical  configurations 
have been examined analytically \cite{Pet:85,Hel.OuY:89,Pet:89,BYW:96}. 
A decomposition of geometries with a convex hull
into spherical harmonics has also proven very effective \cite{Boz.Sve.Zek:97}. 
In the statistical mechanical context, while arbitrary background geometries 
have been considered, the focus has been on the part of the second variation 
contributing at one-loop to the renormalization  of the parameters of the 
theory \cite{For:86,Kle:86,Lei.Pel:87,Dav:89}. 
In any case, there is a gap in perturbation theory which we would like to fill. 
We will write down the second variation of the
hamiltonian about an arbitrary equilibrium configuration,
approaching the problem from a manifestly geometrically covariant point of view.
Two features of the approach we will adopt should be emphasized.
At the level of the second variation, it is no longer justified
to ignore tangential deformations of the membrane. This is because a finite
tangential deformation unlike its infinitesimal counterpart
is not a simple reparametrization of the surface. Indeed, it was by examining
finite tangential deformations of a given surface of constant negative Gaussian curvature 
preserving the curvature that Backlund, Bianchi and others 
generated new surfaces \cite{Rog.Sch:02}. 
We show, however, that when the background geometry is an 
equilibrium, tangential deformations contribute only boundary terms
to the second variation of the Hamiltonian, and therefore for the consideration of 
bulk fluctuations about equilibrium, they can be neglected without incurring any error.
A second simple but subtle technical point is the exploitation of
the fact that the variation  of the divergence of a vector density is equal to the 
divergence of the variation. Though this is not so essential computationally at the 
quadratic level we will work at, by facilitating the
isolation of boundary terms, it does make higher order expansions in the fluctuations about a 
non-trivial background feasible in practice. 

The paper is organized as follows. In section 2, we introduce
our conventions for the geometrical description of a hypersurface
embedded in $R^{N+1}$. We extend our considerations to an $N$-dimensional
hypersurface because of the little extra cost involved and because of potential
applications in other geometrical problems in soft matter physics.
In section 3, we examine how the geometry
changes under an infinitesimal deformation of the hypersurface.
This is used in section 3 to derive the first variation of a geometrical hamiltonian 
for lipid membranes, and to obtain its Euler-Lagrange derivative. In section 4 the
the second variation is examined. In particular, we check that it vanishes at equilibrium
when the deformation corresponds to a rigid normal translation.
We end in section 6 with a few concluding remarks.

\section{Geometry}

We begin by describing briefly the geometry of
a hypersurface  embedded in Euclidean space $R^{N+1}$. 
This allows us to introduce our 
conventions. We will emphasize the peculiarities
associated with a surface embedded in $R^3$.
At the end of this section, we also address the
issue of identifying the
low order independent reparametrization invariants one can construct from the
geometrical quantities that characterize the hypersurface.
 
Consider an orientable hypersurface
$\Sigma$ embedded in $R^{N+1}$. This surface can
be specified locally in parametric form  by $N+1$ shape functions,
\begin{equation}
{\bf x} = {\bf X}(\xi^a )\,,
\end{equation}
where ${\bf x} = x^\mu = (x^1,\cdots, x^{N+1})$ 
are coordinates for $R^{N+1}$,  $\xi^a$ arbitrary
coordinates on the surface $\Sigma$
($a, b, \cdots = 1,\cdots, N$), and
${\bf X}= (X^1,\cdots,X^{N+1})$ are the shape functions.

The Euclidean metric on $R^{N+1}$
induces the metric $g_{ab}$ on $\Sigma$ defined by,
\begin{equation}
g_{ab} := {\bf e}_a \cdot {\bf e}_b \,,
\label{im}
\end{equation}
where the $N$ tangent vectors are defined by
${\bf e}_a (\xi^a ) = \partial_a {\bf X}$ 
($\partial_a := \partial / \partial \xi^a $). 
Latin indices are lowered and raised with $g_{ab}$, and 
its inverse $g^{ab}$, respectively.
The metric $g_{ab}$ determines the intrinsic 
geometry of the hypersurface $\Sigma$. It defines the unique torsionless 
covariant derivative $\nabla_a$ compatible with it, {\it i.e.} satisfying
$\nabla_a g_{bc} = 0$ and
$ ( \nabla_a \nabla_b - \nabla_b \nabla_a ) f (\xi^a ) = 0$
for some surface function $f(\xi^a )$. In terms of the
Christoffel symbol $\Gamma_{ab}^c$,
, acting on a hypersurface vector $V^a$, it reads
\begin{equation}
\nabla_a V^b = \partial_a V^b + \Gamma_{ac}^b V^c \,, 
\end{equation}
where
\begin{equation}
\Gamma_{ab}^c : = g^{cd} {\bf e}_d  \cdot \partial_a {\bf e}_b
= {1 \over 2} g^{cd} \left(
\partial_a g_{bd} + \partial_b
g_{ad} - \partial_d g_{ab} \right)\,.
\label{eq:csy}
\end{equation}
The intrinsic Riemann
curvature ${\cal R}^a{}_{bcd}$ of $\nabla_a$ is defined by
\begin{equation} 
( \nabla_{a} \nabla_{b} - \nabla_b \nabla_a ) V^c 
=: {\cal R}^c{}_{dab} V^d\,.
\end{equation}
In terms of the Christoffel symbol, the Riemann curvature tensor
is given by
\begin{equation}
{\cal R}^a{}_{bcd} =  \partial_c \Gamma_{db}^a -
\partial_d \Gamma_{cb}^a 
+ \Gamma_{ce}^a \Gamma_{db}^e - \Gamma_{de}^a \Gamma_{cb}^e
\,.
\label{eq:curv3}
\end{equation}
The Ricci tensor is defined by contraction,
${\cal R}_{ab} := {\cal R}^{c}{}_{acb}$;
the scalar curvature ${\cal R}$ is defined by
${\cal R} :=  g^{ab} {\cal R}_{ab} $.

When the hypersurface $\Sigma$
is two-dimensional, $N=2$, the Riemann curvature tensor
is completely determined by the scalar curvature:
\begin{equation}
{\cal R}_{abcd} = {{\cal R} \over 2} ( g_{ac}
g_{bd} - g_{ad} g_{bc} )\,.
\label{eq:curv}
\end{equation} 
Note that, as a consequence,
we then have that in two dimensions the Einstein tensor vanishes, 
\begin{equation}
{\cal G}_{ab} := {\cal R}_{ab} - {1 \over 2} {\cal R} g_{ab} =0\,.
\label{eq:curv1}
\end{equation} 
The scalar curvature of a two-dimensional surface is related to 
the Gaussian curvature $G$ of the surface by ${\cal R} = 2 G$.

The simplest geometrical quantity invariant under
reparameterizations of the surface 
one can construct out of the intrinsic geometry
of the hypersurface $\Sigma$ is its area, 
\begin{equation}
A := \int dA = \int d^N \xi 
\; \sqrt{g}\,,
\end{equation}
where $g$ denotes the determinant of the metric $g_{ab}$. 
The next order invariant depending only on the 
intrinsic geometry of the surface is the average scalar curvature over
the surface, $\int dA \; {\cal R}$. For a 
two-dimensional surface with no boundary, by the well-known
Gauss-Bonnet
theorem, this is not only a reparameterization invariant, 
it is also a topological invariant, 
with
\begin{equation}
 \int dA \; {\cal R }  = 4 \pi (1 - {\rm g})\,,
\label{eq:gb}
\end{equation}
where ${\rm g}$ is the genus of the surface.

Let us turn now to the extrinsic geometry of $\Sigma$.
The single normal vector ${\bf n} (\xi^a ) $ to $\Sigma$ 
in $R^N$  can be defined in implicit form by, 
\begin{equation}
 {\bf e}_a\, \cdot {\bf n}   = 0\,,
\label{eq:dn1}
\end{equation}
with the normalization,
 \begin{equation}
{\bf n}\cdot {\bf n} = 1\,.
\label{eq:dn2}
\end{equation} 
Note that these equations determine ${\bf n}$ only up to a sign:
the normal can be inward or outward. Since the surface
is assumed to be orientable, we can pick one sign consistently,
and we choose the normal to be outward. 
The unit normal vector field can also be given explicitly as
\begin{equation}
n_\mu =  {1 \over N! \sqrt{g}} \;
\varepsilon_{\mu \rho_1 \cdots \rho_{N-1}} \; \varepsilon^{a_1\cdots
a_{N-1}} \; 
(e^{\rho_1}{}_{a_1} ) \cdots (e^{\rho_{N-1}}{}_{a_{N-1}} )\,,
\label{eq:n2}
\end{equation}
where $ \varepsilon_{\mu_1\cdots \mu_N}$ and $\varepsilon^{a_1\cdots 
a_{N-1}}$ are 
the totally antisymmetric Levi-Civita symbols for $R^{N+1}$
and $\Sigma$, respectively ($ \varepsilon_{1\cdots N} = + 1$). 
The factor of $\sqrt{g}$
is necessary in order to make ${\bf n}$ a scalar under
reparametrizations. 

When $\Sigma$ is  closed, as we are assuming in this paper, 
the total volume enclosed
by the surface $\Sigma$ in $R^{N+1}$ is an invariant.
The normal to $\Sigma$, as given by (\ref{eq:n2}), 
 allows us to provide an alternative
definition of the volume occupied by the interior of 
$\Sigma$ as the surface integral
\begin{equation}
V = {1 \over N+1} \int dA \; {\bf n}\cdot {\bf X}\,.
\label{eq:vola}
\end{equation}
 
The space vectors $\{ {\bf n}, {\bf e}_a \}$
form a basis adapted to the hypersurface. Their gradients
along the surface are themselves space vectors, and can be 
decomposed in turn with respect to this basis. These
decompositions constitute the classical 
Gauss-Weingarten equations,
\begin{eqnarray}
\partial_a {\bf e}_b  &=& 
\Gamma_{ab}^c \; {\bf e}_c -
K_{ab} \; {\bf n}\,, \label{eq:gw1} \\
\partial_a {\bf n} &=& K_{ab} \; g^{bc} \; {\bf e}_c\,.
\label{eq:gw2}
\end{eqnarray}
Here $\Gamma_{ab}^c$ is the Christoffel symbol
defined
in  (\ref{eq:csy}).
The extrinsic curvature of $\Sigma$ is given by the symmetric rank two 
surface tensor, 
\begin{equation}
K_{ab} := -  {\bf n}\cdot  \partial_a {\bf e}_b= K_{ba}\,.
\end{equation}
(Note that many authors differ by a sign in this definition.) 
We define its trace with respect
to the intrinsic metric,
\begin{equation}
K := g^{ab} K_{ab}=: 2 H\,,
\end{equation}
where $H$ represents the mean extrinsic curvature of the surface.

The intrinsic and the extrinsic  geometries of 
$\Sigma$, determined respectively by $g_{ab}$ and by $K_{ab}$,
cannot be specified independently. They
are related by the well-known integrability conditions of 
Gauss-Codazzi, and Codazzi-Mainardi, given respectively by,
\begin{eqnarray}
{\cal R}_{abcd} - K_{ac} K_{bd} + K_{ad} K_{bc} &=& 0 \,, 
\label{eq:gauss}\\
\nabla_a K_{bc} - \nabla_b K_{ac} &=& 0 \,.
\label{eq:cm}
\end{eqnarray}
These equations follow as integrability conditions 
from taking a gradient along the surface of
of the Gauss-Weingarten
equations  (\ref{eq:gw1}), (\ref{eq:gw2}), and
then the appropriate anti-symmetric part. 

The fundamental theorem for surfaces states 
that, given $g_{ab}$ and 
$K_{ab}$, these equations are not only necessary,
but also sufficient for the existence of
an embedding with these quantities as intrinsic
metric and extrinsic curvature. Furthermore, the
embedding is unique, up to rigid motions in the
ambient space (see {\it e.g.} \cite{Spivak}). 

Contraction of the
Gauss-Codazzi-Mainardi equations 
with the contravariant intrinsic metric $g^{ab}$ results in 
\begin{eqnarray}
 {\cal R}_{ab} -  K K_{ab} +  K_{ac} K_b{}^c &=& 0 \,, 
\label{eq:gausscc}\\
{\cal R} - K^2 + K_{ab} K^{ab} &=& 0 \,, 
\label{eq:gaussc}\\
\nabla_a K_b{}^a - \nabla_b K  &=& 0\,.
\label{eq:cmc}
\end{eqnarray}
For a two-dimensional surface,
the contracted equations  (\ref{eq:gaussc}) and (\ref{eq:cmc}) possess 
the same content 
as the full Gauss-Codazzi-Mainardi equations, (\ref{eq:gauss}) and (\ref{eq:cm}).

The lowest order geometrical invariant 
which involves the extrinsic geometry of the 
surface is the 
trace of the extrinsic curvature integrated  over of the surface,
\begin{equation}
M := \int dA \; K\,.
\end{equation}
This invariant, linear in ${\bf n}$, depends
on the orientation of the surface. Note that
$[ K ] = L^{-1} $, so that $[M] = L $.
The next order reparameterization
invariants are quadratic in the extrinsic curvature ($\approx L^0$ ),
\[
\int dA \; K^2\,,   \quad 
\int dA \;  K^{ab} K_{ab}\,,
\]
together with the Gaussian term,
$\int dA \; {\cal R}$.
In general, these three invariants  are not independent, as follows 
from the contracted Gauss equation,
(\ref{eq:gaussc}).
Furthermore, in two dimensions,  
the Gaussian term is the Gauss-Bonnet topological invariant,
see (\ref{eq:gb}).
For a two-dimensional surface,  at this order
there is then only one independent local
scalar associated with the embedding, and it is customary to choose
$\int dA \; K^2$. 
The gradients of the extrinsic curvature $\nabla_a K_{ab}$ which 
appear in the  Codazzi-Mainardi equations do not feature. 
Differential bending  can be ignored up to this order.

Let us now construct explicitly the geometrically independent terms 
of order $L^{-2}$ and lower. 
In general,  at order $L^{-1}$, we have three independent scalars.
 Of the set,
$K^3, K K^{ab} K_{ab}, K^{ab} K_{bc} K^{c}{}_a$, 
$K {\cal R}$, and $K_{ab} {\cal G}^{ab}$, 
the Gauss-Codazzi equations can be exploited to produce
three that are independent
\[
\int dA \; K^3  \,, \quad \int dA \; K \,{\cal R} \,,
 \quad {\rm and}\quad 
\int dA \;  K^{ab} {\cal G}_{ab}\,.
\]
The last of the three vanishes identically for two dimensional surfaces.

At  order $L^{-2}$  
using the Gauss-Codazzi equation (\ref{eq:gauss}), it is easy to show
that the independent scalars are
${\cal R}^{abcd} {\cal R}_{abcd}$, ${\cal R}^{ab}
{\cal R}_{ab}$, ${\cal R}^2$, $K^4$, ${\cal R} K^2$, and ${\cal G}_{ab}
K^{ab}$. Of these, for a two-dimensional surface, only 
${\cal R}^2$, $K^4$, ${\cal R} K^2$ survive.
Moreover, at this order
gradients of the extrinsic geometry enter.
 We can reduce all terms quadratic in $\nabla_a K_{bc}$ to the 
form $\nabla_a K \nabla^a K$ plus 
terms that we have already considered. 
We exploit the uncontracted
Codazzi-Mainardi integrability 
condition (\ref{eq:cm}) to write 
\[
\int dA\; (\nabla_a K_{bc}) (\nabla^a K^{bc} ) =
\int dA \; (\nabla_a K_{bc})(\nabla_b K_{ac}) 
= - \int dA \; K_{bc} \nabla_a\nabla_b K^{ac} \,,
\]
where we have integrated by parts.
We now note that
\[ 
[\nabla_a,\nabla_b] K^{ac}= {\cal R}^a{}_{dab} K^{dc}
+ {\cal R}^c{}_{dab} K^{ad}\,,
\]
so that
\begin{eqnarray}
\int dA \; (\nabla_a K_{bc}) (\nabla^a K^{bc} ) &=& -
\int dA  
K_{c}{}^b
( \nabla_b \nabla_a K^{ac} +
{\cal R}_{db} K^{dc}
+ {\cal R}^c{}_{dab} K^{ad}) \nonumber
\\
&=& \int dA \left( K^{bc} \nabla_b \nabla_c K
+{\cal R}_{db} K^{dc} K^b{}_c
+ {\cal R}_{cdab} K^{ad} K^{bc}
\right)\,, \nonumber
\end{eqnarray}
where we have used the contracted Codazzi-Mainardi equation 
(\ref{eq:cmc}) in the first term. We now integrate by parts 
again the first term, and use the Gauss-Codazzi-Mainardi equations
 to
obtain
\[
\int dA \; (\nabla_a K_{bc}) (\nabla^a K^{bc} ) 
=
\int dA  
(\nabla^a K) (\nabla_a K ) 
+ {\cal R}^{ab} {\cal R}_{ab} - K K^{ab} {\cal R}_{ab}
+ {1 \over 2} {\cal R}_{abcd} {\cal R}^{abcd} 
\,.
\]
We conclude that for a two-dimensional surface, at this order there 
are four independent scalars, 
\[
\int dA \;\left[ {\cal R}^2\,, \quad \int dA \; {\cal R} K^2\,,
\quad \int dA \; K^4\,,  \quad {\rm and}\quad 
\int dA \; (\nabla^a K)( \nabla_a K) \right]\,.
\]
These higher order terms appear in geometric models for the so called egg-carton membranes
\cite{GH96}, and for tubular structures \cite{FG}.
Note that if $K$ is treated as a simple 
scalar field, $\phi$, then the model described at this order
is a $\lambda\phi^4$ theory non-minimally coupled to the intrinsic curvature. 
The Euler-Lagrange equations of the two models are, of course, different.

\section{Deformations}

In this section, we consider infinitesimal deformations 
of a hypersurface $\Sigma$, and how the various geometrical
quantities that characterize it change under
deformation. 
A one-parameter  deformation of the hypersurface is described by the 
functions ${\bf X}(\xi^a,u)$. 
An infinitesimal change of the embedding functions 
\begin{equation}
{\bf X} \to {\bf X} (\xi) + \delta {\bf X} (\xi)\,,
\end{equation}
is characterized by the infinitesimal vector $\delta {\bf X} (\xi) 
= \partial_u {\bf X}(\xi^a,u)|_{u=0} \delta u $.
This vector can be decomposed
into its components normal and tangential to the hypersurface $\Sigma$: 
\begin{equation}
\delta {\bf X} =
\delta_\perp {\bf X} + \delta_\parallel {\bf X}
= \Phi {\bf n} + \Phi^a {\bf e}_a \,.
\label{eq:def}
\end{equation}
The components $\Phi, \Phi^a$ have the dimensions of length.
We will assume that they are much smaller than any
characteristic length in the system, such as the curvature radii.

We first determine  how the basis $\{ {\bf e}_a , {\bf n} \} $ 
changes under deformation. 
The key observation is that 
$\partial_u$ and $\partial_a$ commute, so that 
the change in the tangent vector satisfies 
\begin{equation}\delta {\bf e}_a  =
\partial_a  ( \delta{\bf X}) \,.\end{equation}
We now  decompose $\delta {\bf X}$ into its tangential 
and normal parts according to 
(\ref{eq:def}) so that 
\begin{equation}
\delta {\bf e}_a  
= (\nabla_a \Phi^b) {\bf e}_b 
- K_{ab} \Phi^b {\bf n} + (\nabla_a \Phi ) {\bf n}
+ \Phi K_{ab} g^{bc} {\bf e}_c 
\label{eq:dt}
\,,
\end{equation}
where we have used the Gauss-Weingarten equations (\ref{eq:gw1}) 
and (\ref{eq:gw2}).
We thus have for the deformations induced by $\delta{\bf X}_{||}$ and 
$\delta{\bf X}_\perp$, respectively,
\begin{equation}
\delta_{\parallel} {\bf e}_a = (\nabla_a \Phi^b) {\bf e}_b 
- K_{ab} \Phi^b {\bf n} \,, \quad \quad  
\delta_{\perp} {\bf e}_a =  (\nabla_a \Phi ) {\bf n}
+ \Phi K_{ab} g^{bc} {\bf e}_c \label{eq:dt1}\,.
\end{equation}
For the deformation of the induced metric, it follows that
\begin{equation}
\delta_\parallel g_{ab} 
=
\nabla_a \Phi_b + \nabla_b \Phi_a \,,
\quad \quad
\delta_\perp g_{ab} 
= 2 K_{ab} \Phi\,.
\end{equation}
The tangential deformation is just the 
the Lie derivative along the surface vector field $\Phi^a$.
The normal deformation summarizes the geometrical content
of the extrinsic curvature as one half of 
Lie derivative
of the intrinsic metric along the normal vector field.
Note that for the inverse
metric one has that
$\delta_\perp g^{ab} = - 2 K^{ab} \Phi$.
It follows from these relations that
the first order deformation of the infinitesimal
area element is 
\begin{equation}
\delta_\parallel dA =  dA \; \nabla_a \Phi^a\,,
\quad \quad
\delta_\perp dA = dA \; K \; \Phi\,.
\label{eq:sqrt}
\end{equation}
This last expression encodes the geometrical content
of the trace of the extrinsic  curvature as the relative change of 
area per unit normal deformation.

The variations of the normal vector, 
\begin{equation} 
\delta_{\parallel} {\bf n} =  K_{ab} \; \Phi^a \; g^{bc} {\bf e}_c\,,
\quad \quad
\delta_\perp {\bf n} = - (\nabla_a \Phi ) g^{ab} {\bf e}_b \,,
\label{eq:defn}
\end{equation}
follow readily from the defining relations
(\ref{eq:dn1}) and (\ref{eq:dn2}), together with 
(\ref{eq:dt1}). 

For the remainder of this section we will focus on normal
deformations of the hypersurface. The reason is that 
infinitesimal tangential deformations correspond to reparametrizations
of the surface, and, as shown in the next section, they
contribute only boundary terms to the variation of global
quantities.

It is possible to evaluate the 
first order normal variation of the intrinsic
scalar curvature ${\cal R}$ either intrinsically or extrinsically,
the latter using the 
Gauss-Codazzi equation. Let us consider the former approach. 
Using the definition of the
Riemann tensor given by  (\ref{eq:curv3}), we have that
the variation of the Riemann tensor is (see {\it e.g.} \cite{Wald})
\[
\delta {\cal R}^a{}_{bcd} =  \nabla_c (\delta \Gamma_{db}^a
)
 - \nabla_d (\delta \Gamma_{cb}^a )\,,
\]
which implies 
\[
\delta {\cal R}_{ab} =  \nabla_c (\delta \Gamma_{ab}^c )
 - \nabla_b (\delta \Gamma_{ca}^c )\,.
\]
For the scalar curvature we have then that its normal variation is
\begin{eqnarray}
\delta_\perp {\cal R} &=& (\delta_\perp g^{ab} )
 {\cal R}_{ab} + g ^{ab} (\delta_\perp {\cal R}_{ab} )
\nonumber \\ 
&=& - 2{\cal R}_{ab} K^{ab} \Phi 
 + \nabla_c ( g^{ab} \delta_\perp \Gamma_{ab}^c )
 - \nabla^a (\delta_\perp \Gamma_{ca}^c )\,.
\nonumber 
\end{eqnarray}
Now, an arbitrary variation of the metric induces a 
variation 
\[
\delta \Gamma_{ab}^c = {1\over 2} g^{cd} (\nabla_b
\delta g_{ad} +
\nabla_a \delta g_{bd} -\nabla_d \delta g_{ab})
\]
in the Christoffel symbols,
so that using the 
Codazzi-Mainardi equations,  we have
that the normal variation of the Christoffel symbol is
\begin{equation}
\delta_\perp \Gamma_{ab}^c =
K_a{}^c \nabla_b \Phi + K_b{}^c \nabla_a \Phi
- K_{ab} \nabla^c \Phi
+ (\nabla_a K_b{}^c ) \Phi
\,.\label{eq:delgam}
\end{equation}
As a consequence of the Codazzi-Mainardi
equations, the (apparently unsymmetric)
last term on the right hand side is, in fact, 
symmetric under the interchange of tangent
indices.
We determine then that the scalar curvature 
varies according  to
\begin{equation}
\delta_\perp {\cal R} = 
- 2{\cal R}_{ab} K^{ab} \Phi 
+ 2 \nabla_a \left[ ( K^{ab}- g^{ab} K)
\nabla_b \Phi \right]\,.
\label{eq:vcurv1}
\end{equation}
Note that for the densitized scalar curvature we have
\begin{equation}
\delta_\perp \sqrt{g} {\cal R} = 
\sqrt{g} ( - 2{\cal G}_{ab} K^{ab} \Phi
+ 2 \nabla_a \left[ ( K^{ab}- g^{ab} K)
\nabla_b \Phi \right]\,.
\label{eq:vcurv}
\end{equation}
It follows that
the variation of the integrated scalar curvature is a pure divergence
in two dimensions, as expected from the Gauss-Bonnet theorem
(\ref{eq:gb}).

Let us consider now the first order normal deformation of the
extrinsic geometry. 
We have that 
\begin{eqnarray}
\delta_\perp K_{ab} &=& - (\delta_\perp {\bf n})\cdot \partial_a
{\bf e}_b  - {\bf n}\cdot \partial_a ( \delta_\perp {\bf e}_b 
) \nonumber \\ 
&=&
(\nabla_c \Phi ) {\bf e}^c \cdot \partial_a
{\bf e}_b - {\bf n} \cdot \partial_a [(\nabla_b \Phi) {\bf n}
+ \Phi K_{bc} {\bf e}^c ]
\nonumber \\ &=&
 \Gamma_{ab}^c \nabla_c \Phi 
- \partial_a \nabla_b \Phi +  K_{ac} K^c{}_b \Phi
\nonumber \\ &=&
- \nabla_a \nabla_b \Phi +  K_{ac} K^c{}_b \Phi\,.
\end{eqnarray}
This is a remarkably simple expression.
Using the contracted Gauss equation (\ref{eq:gausscc}) we can cast
this expression in the alternative form, 
\begin{equation}
\delta_\perp K_{ab} 
= - \nabla_a \nabla_b \Phi + (K K_{ab}
- {\cal R}_{ab} ) \Phi\,.
\label{eq:Kab1}
\end{equation}
For the trace of the  extrinsic curvature 
this gives (recall that $\delta_\perp g^{ab} = - 2 K^{ab} \Phi$),
\begin{equation}
\delta_\perp K = - \Delta \Phi + ( {\cal R} - K^2 ) \Phi \,,
\label{eq:dpc}
 \end{equation}
where $\Delta := g^{ab} \nabla_a \nabla_b$ denotes the laplacian on
$\Sigma$, and we have used the contracted Gauss equation
(\ref{eq:gaussc}). 

As a check of the consistency of the expressions we have derived, one 
can verify that they imply 
the vanishing of the first order variation
of the contracted Gauss equation,  {\it i.e.}
$\delta_\perp ( {\cal R} - K^2 + K_{ab} K^{ab}) = 0$.

In the calculations of higher order
variations, an essential ingredient
is the commutator of the
deformation operator and the covariant derivative on $\Sigma$. 
This commutator allows 
one to express
deformations of the covariant derivative of any geometrical
quantity of interest in terms of the covariant derivative
of the variation of such quantity. For instance, when
 acting on an arbitrary second rank tensor $A_a{}^b$ in $\Sigma$, it is
given by 
\begin{equation}
[ \delta_\perp , \nabla_a ] A_b{}^c = - (\delta_\perp
\Gamma_{ab}^d ) \; A_d{}^c
+ (\delta_\perp \Gamma_{ad}^c )\; A_b{}^d \,,
\end{equation}
where $\delta_\perp \Gamma_{ab}^c$ is defined in (\ref{eq:delgam}).
Also for this case, it is a useful check of the validity
of these expressions, to verify that 
the first order variation of the Codazzi-Mainardi equation
(\ref{eq:cm}), which involves such commutator,
vanishes identically.

A useful expression is the commutator of the deformation derivative
with the Laplacian, acting on an arbitrary function $f$,
\begin{equation}
[ \delta_\perp , \Delta  ] f = 
 - 2 K^{ab} \Phi \nabla_a \nabla_b f  
- 2  K^{ab}   \nabla_a f \nabla_b  \Phi
+  K \nabla^a f \nabla_a \Phi -  (\nabla_a K) (\nabla^a f ) \Phi\,.
\label{eq:lapla}
\end{equation}

\section{First variation}

In this section, we apply the formalism for deformations
we have developed  to the systematic 
determination of the Euler-Lagrange derivatives of the geometrical
invariants 
which appear in the  hamiltonian for the strict bilayer couple model  
\begin{equation}
F [ X] = \alpha \int dA\; K^2  + \beta M  + \mu A  + P V\,.
\label{eq:hami}
\end{equation}
The first term is the bending energy, with $\alpha$ the bending rigidity,
the constants  $ \mu, P, \beta$ are lagrange multipliers that
enforce the constraints of constant area, constant volume and constant
area difference, respectively. For the sake of simplicity, we have 
not included the non-local bending rigidity term, necessary in a realistic
description of lipid vesicles \cite{BSZ,Miao}.

We first comment on tangential deformations.
For the surface area element we have that under a tangential deormation
it transforms according to (\ref{eq:sqrt}).
In addition, any surface scalar satisfies 
$\delta_\parallel f (X) = \Phi^a \partial_a f (X)$.
Thus, any reparameterization invariant
functional of the form
\begin{equation}
F [X] = \int dA f (X)\,,
\label{eq:Ff}
\end{equation}
deforms tangentially as
\begin{equation}
\delta_\parallel F [X] = \int dA \; \nabla_a [ \Phi^a f (X) ]\,.
\end{equation} 
Using Stokes theorem this becomes
\begin{equation}
\delta_\parallel F [X] = \int
 ds \; \eta_a \; \Phi^a \; f (X) \,,
\label{eq:bdry}
\end{equation}
where the integral ranges over the boundary of $\Sigma$ and $\eta^a$ 
denotes the unit normal to the boundary into $\Sigma$.
If this boundary is empty, the integral vanishes identically.
It is worthwhile emphasizing that in the case of a 
surface with a boundary, 
this integral in general will
be non-vanishing, and it is no longer correct
to neglect the tangential variations of the 
surface. In fact, in the variational principle,
the vanishing of such terms will determine 
the boundary conditions to be imposed on the shape functions
(see {\it e.g.} \cite{Cap.Guv.San:02a}).

If the surface has no boundary, the tangential part of the variation 
can always be associated with a reparametrization. 
Since we are interested, to begin with,
in quantities that are invariant under
surface reparametrizations, we disregard
this contribution and focus on normal
deformations of the 
geometry of $\Sigma$. We write
\begin{equation}
\delta_\perp F [ X ] = \int dA \; \left\{ {\cal E}[f] \; \Phi  + 
\nabla_a V^a_{(1)} [f] \right\}\,.
\label{eq:Ffv}
\end{equation}
The divergence comes about when we integrate by parts to 
remove all derivatives from the normal deformation $\Phi$.
The subscript in the vector appearing in the second term
refers to the order of the variation.

We begin with the first order normal variation of $A$ and $V$.
For the normal variation of the total area it follows immediately from
(\ref{eq:sqrt})
that
\begin{equation}
\delta_\perp A  = \int dA \; K \; \Phi \,.
\label{eq:d1A}
\end{equation}
The difference in area between two surfaces 
separated by an infinitesimal constant 
normal  distance $\Phi$ is proportional to the integrated mean curvature.
The area of the surface is extremal for arbitrary normal 
deformations, $\delta_\perp A  =0$, when $K=0$
at each point on the surface. 

One way to determine the normal deformation of 
the volume enclosed by the surface
is to exploit the definition given in (\ref{eq:vola}). We have 
\begin{eqnarray}
\delta_\perp V &=& {1 \over N+1} \int d^N \xi \;
[ (\delta_\perp \sqrt{g} )\;  {\bf n}\cdot {\bf X} + \sqrt{g} \; (\delta_\perp
{\bf n} ) \cdot {\bf X} 
+ \sqrt{g} \; {\bf n}\cdot  \delta_\perp {\bf X} ]
\nonumber \\
&=& {1 \over N+1} \int dA \; [ \Phi\;  K \; {\bf n} \cdot {\bf X}
- (\nabla_a \Phi ) \; g^{ab} \; {\bf e}_b \cdot {\bf X}
+ \Phi ]\,.
\end{eqnarray}
Integrating by parts the second term and dropping
a total divergence gives
\begin{eqnarray}
\delta_\perp V &=& {1 \over N+1} \int dA \; [ \Phi\;  K \; {\bf n}\cdot
{\bf X}
+ \Phi \;  g^{ab} \; \partial_a {\bf e}_b 
\cdot {\bf X} + \Phi \; g^{ab} \; {\bf e}_b\cdot {\bf e}_a 
+ \Phi ]\nonumber \\
&=& {1 \over N+1} \int_\Sigma dA \; [ \Phi \; K \; {\bf n} \cdot {\bf X}
- \Phi  \; K \; {\bf n}\cdot
{\bf X} + N \Phi + \Phi ]\nonumber \\
&=& \int dA\; \Phi \,,
\label{eq:d1V}
\end{eqnarray}
where we have used the Gauss-Weingarten equations (\ref{eq:gw1}) to obtain  the
second term of the second line.
This final expression should come as no surprise. The infinitesimal change in 
volume is simply proportional to the area of the  surface times the
normal displacement.

We now examine the integrated powers of the mean extrinsic curvature.
The deformation of the density $\sqrt{g} K$ , using  (\ref{eq:sqrt}),
 (\ref{eq:dpc}), is 
\begin{equation}
\delta_\perp \sqrt{g} \; K =  - \Delta \Phi + {\cal R} \; \Phi \,.
\label{eq:d1m}
\end{equation}
We thus find for the total mean curvature, up to a total
divergence,
\begin{equation}
\delta_\perp M =  \int dA \;  {\cal R} \; \Phi \,.
\label{eq:aK1}
\end{equation}
It is interesting that this expression 
depends only on the intrinsic geometry of the surface.
If the scalar curvature vanishes, 
so also does $\delta_\perp M $. This is to be expected in
a two-dimensional surface where ${\cal R}=0$ implies that it is  flat.  
It is, however, a non-trivial statement for higher dimensions.

For the variation of the integrated second power of the extrinsic curvature  
we have
\begin{equation}
\delta_\perp  \int dA \; K^2  
  =  \int dA \; K \; [ - 2  \Delta  \Phi
+ ( 2 {\cal R} - K^2) \Phi ] \,;
\label{eq:m2K1}
\end{equation}
We now integrate by parts twice to obtain, again up to a total divergence,
\begin{equation}
\delta_\perp  \int dA \; K^2  
=  \int dA \;  \left[ - 2  \Delta  K
+ ( 2 {\cal R} - K^2) K \right] \Phi\,.
\label{eq:m2K1b}
\end{equation}

We are now in the position to derive the Euler-Lagrange
derivative for the model defined by the Helfrich hamiltonian 
(\ref{eq:hami}).
Using the 
expressions (\ref{eq:d1A}), (\ref{eq:d1V}), (\ref{eq:aK1}),
(\ref{eq:m2K1b}), in (\ref{eq:Ffv}), we find \cite{Hel.OuY:87} 
\begin{equation}
{\cal E}   = - 2 
\alpha \Delta  K
 +   (2 \alpha K + \beta){\cal R} 
-  \alpha K^3 + \lambda K + P  \,.
\label{eq:shape}
\end{equation}
At equilibrium, ${\cal E} = 0$, it is  known as the shape equation.
Note that it involves four derivatives of the shape functions.
It determines the equilibrium configurations of lipid membranes 
described by the hamiltonian (\ref{eq:hami}).
Recently it was shown how to cast the shape equation in the form of a conservation law
\cite{Cap.Guv:02}.

\section{Second variation}

In this section, we derive the second variation of the 
hamiltonian (\ref{eq:hami}). 
Before we proceed, we note that because the shape functions
${\bf X}$ are  the variables to be varied,
we have that
$\delta_\perp \Phi \ne  0\,,$
to second order, and similarly for the remaining three possibilities, 
$\delta_\perp \Phi^a$, $\delta_{||} \Phi$ and $\delta_{||} \Phi^a$.
In fact, using (\ref{eq:dt1}), (\ref{eq:defn}), we have
\begin{equation}
\delta_\perp \Phi = - \Phi^a \nabla_a \Phi\,,
 \quad 
\delta_\parallel \Phi = K_{ab} \Phi^a \Phi^b\,,
\end{equation}
together with
\begin{equation}
\delta_\perp \Phi^a = \Phi \nabla^a \Phi + K^a{}_b \Phi^b \Phi\,, \quad 
\delta_\parallel \Phi^a = \Phi^b \nabla^a \Phi_b - K^a{}_b \Phi^b \Phi\,.  
\end{equation}
This appears to suggest that the computation of the second variation is a
formidable task and that the decompostion of deformations into tangential and 
normal parts will not be so useful. As we will show, the formal apparatus 
we have developed simplifies the calculation enormously and all of the troublesome 
tangential terms are collected in a divergence when the 
Euler-Lagrange equations are satisfied: no actual error 
is incurred in setting
$\delta_\perp \Phi=0$ and neglecting $\Phi^a$ in the sequel. 

Let ${\cal F}=\sqrt{g} f$, where $f$ is a scalar, introduced in (\ref{eq:Ff}).
At first order, we have
\begin{equation}
\delta F = \int d^N\xi \; \left[ \delta_{||} {\cal F} + 
\delta_\perp {\cal F} \right] = 
\int d^n\xi \; \left[\sqrt{g} {\cal E}\Phi + 
\nabla_a ({\cal V}_{(1)}[{\cal F}] + {\cal F}\Phi^a)\right]\,,
\end{equation}
where ${\cal V}_{(1)}^a [{\cal F}]= \sqrt{g} V_{(1)}^a[f]$ is the densitized vector defined by
Eq.(\ref{eq:Ffv}).
Now the divergence of a vector
density is independent of the surface affine connection
$\Gamma_{ab}^c$,  and thus 
variation of the divergence of a vector density ${\cal V}^a$
is equal to the divergence of the variation,
\begin{equation}
\delta  (\nabla_a {\cal V}^a)
= \nabla_a (\delta {\cal V}^a)\,.
\end{equation}
We thus have at second order, 
\begin{equation}
\delta^2 F = 
\int d^N\xi \; \delta \left[\sqrt{g} {\cal E}\Phi\right]
\,,
\end{equation}
modulo another divergence. Thus, whereas for the second variation of ${\cal F}$ we 
cannot legitimately discard the divergence, we can for $F$. Furthermore, we have
\begin{equation}
\delta \left[\sqrt{g} {\cal E}\Phi\right] = 
\delta_\perp (\sqrt{g} {\cal E}) \Phi 
+ \nabla_a (\sqrt{g} {\cal E}\Phi^a)\Phi + \sqrt{g}{\cal E}\delta\Phi\,.
\end{equation}
Thus, modulo the Euler-Lagrange equation, $\delta \left[\sqrt{g} {\cal E}\Phi\right] = 
\delta_\perp (\sqrt{g} {\cal E}) \Phi$. But $\delta_\perp (\sqrt{g} {\cal E})$ can always be 
expressed as ${\cal L}\Phi$,
for some local differential operator ${\cal L}$, and therefore
the second order variation of the hamiltonian
can itself be written in the form
\begin{equation}
\delta^2 F [X] = \int dA  \;
\Phi \; {\cal L }  \; \Phi \,.
\label{eq:op}
\end{equation}
It must be remembered, however, that this is not true for the individual 
terms contributing to $F$, for which ${\cal E}\ne 0$.
When the second variation of these terms is written down it, it should be understand 
that the expression refers to the second normal variation; off
diagonal terms $\sim \Phi \Phi_a $ as well as purely tangential terms $\sim \Phi_a \Phi_b$
have been suppressed. 

A second order variation can also be seen
as a different deformation
$ X^\mu \to X^\mu (\xi) + \Phi' {\bf n} 
$ applied to the first order variations, and then letting
$\Phi' = \Phi $. From a computational point of view,
this is clearly equivalent
to a repeated application of the normal 
deformation operator $\delta_\perp $.

For the second order variation of the enclosed volume  
no work is needed, since, using (\ref{eq:sqrt}), it is
simply related to the first variation of the area with
\begin{equation}
\delta^2 V = \delta \int dA  \; \Phi =
\int dA \; K \; \Phi^2\,.
\end{equation}

In order to derive the second variation of the area,
various strategies are possible. On one hand one can 
simply
calculate the second variation of the metric.
Using  (\ref{eq:Kab1}) 
one obtains,
\begin{equation}
\delta_\perp^2 g_{ab} 
= 2 \left( \delta_\perp K_{ab} \right) \; \Phi 
=  - 2 \Phi \;[  
\nabla_a \nabla_b \Phi -  K_a{}^c K_{cb}  ]\; \Phi\,,
\end{equation}
so for the area element this gives
\begin{equation}
\delta_\perp^2 \sqrt{g} =
 \sqrt{g} ( - \Phi \Delta  \Phi + {\cal R } \Phi^2 )\,,
\label{eq:2varea}
\end{equation}
and the second variation of the area
is given by the well-known expression
(see {\it e.g.} \cite{Lawson})
\begin{equation}
\delta^2 \, A = \int dA \;
   \Phi \left\{- \Delta   + {\cal R }
\right\}\Phi \,.
\end{equation}

There is an alternative route to the
second variation of the area, which becomes 
increasingly useful as one considers higher variations. The idea
is to exploit the natural hierarchy in the variations
illustrated by the relations
\begin{eqnarray}
\delta_\perp \sqrt{g} &=& \sqrt{g} \; K \; \Phi \nonumber \\
\delta_\perp^2 \sqrt{g} &=& \delta_\perp (\sqrt{g} K ) \Phi\,.
\nonumber
\end{eqnarray}
Using (\ref{eq:d1m}), this gives directly (\ref{eq:2varea}).
The usefulness of this approach is apparent: the
$n^{\rm th}$ variation of the surface volume element
can be expressed in terms of the 
$(n-1)^{\rm th}$ variation of the densitized mean
extrinsic curvature.

We now exploit this observation to derive the
second variation of the densitized mean extrinsic curvature.  
We rewrite (\ref{eq:aK1})  as
\begin{equation}
\delta_\perp (\sqrt{g} K ) = \sqrt{g} \; {\cal R} \;
\Phi + \nabla_a {\cal V}^a_{(1)} [K]\,,
\label{eq:aK1a}
\end{equation}
where we have defined the vector density 
${\cal V}^a_{(1)} [K] :=   - \sqrt{g} \nabla^a \Phi$.
Using the fact that the 
variation of the divergence of a vector density
is equal to the divergence of the variation, we have
\begin{eqnarray}
\delta_\perp^2 (\sqrt{g} K ) &=&
 \delta_\perp \{ \sqrt{g} {\cal R}
\Phi + \nabla_a {\cal V}^a_{(1)} [K] \}
\nonumber \\
&=&  \Phi \delta_\perp ( \sqrt{g} {\cal R}) 
 + \nabla_a {\cal V}^a_{(2)} [K]  \nonumber \\
&=& 2 \sqrt{g} \Phi \left[ 
( K^{ab} - K g^{ab} ) \nabla_a \nabla_b \Phi
-  {\cal G}_{ab} K^{ab} \Phi \right]  
+ \nabla_a  {\cal V}^a_{(2)} [K])\,,
\label{eq:d2aKa}
\end{eqnarray}
where we have used  (\ref{eq:vcurv}), 
and defined
the vector density 
\begin{equation}
{\cal V}^a_{(2)} [K] = \delta_\perp {\cal V}^a_{(1)} [K]
=  \sqrt{g} \Phi ( 2 K^{ab} - g^{ab} K )
\nabla_b \Phi\,.  
\label{eq:va2}
\end{equation} 
We have therefore for the second variation of $M$,
\begin{equation}
\delta^2 M   =  \int dA \; \{ 
2 \Phi \; [ ( K^{ab} - K g^{ab} ) \nabla_a \nabla_b  - 
{\cal G}_{ab} K^{ab} ] \;\Phi \} \,.
\end{equation}
Note that if the surface is flat then the second order variation, 
like the first, vanishes.
Thus, in  a gaussian approximation about a flat background geometry, 
this term is absent.

Our final  task in this section is to compute  
the second variation of the densitized squared
mean extrinsic curvature. 
Unfortunately, this quantity does not fit naturally
in the hierarchy
of variations we have been exploiting. One approach is to
calculate the second
variation of $K$ directly. 
From (\ref{eq:dpc}), using (\ref{eq:vcurv1}), (\ref{eq:lapla}),
we have that
\begin{equation}
\delta_\perp^2 K = 4 K^{ab} \Phi \nabla_a \nabla_b \Phi 
+ (2 K^{ab} - K g^{ab} ) \nabla_a \Phi \nabla_b \Phi
+ 2 K (K^2 - {\cal R} ) \Phi^2 - 2 {\cal R}_{ab} K^{ab} \Phi^2\,.
\end{equation}
We prefer to capitalize 
on the relative simplicity of the expressions
for the variation of the densitized mean curvature
and use the identity
\begin{equation}
\delta_\perp^2 (\sqrt{g} K^2 ) =
2K \;\delta_\perp^2 (\sqrt{g} K ) + 2 [ \delta_\perp (\sqrt{g} K) ]^2
g^{-1/2} - 5 K^2 \Phi \delta_\perp (\sqrt{g} K)
+ 2 \sqrt{g} K^4 \Phi^2\,.
\end{equation}
A short calculation gives 
\begin{eqnarray}
\delta_\perp^2 (\sqrt{g} K^2 ) &=&
\sqrt{g} [ 2 (\Delta  \Phi )^2 
+ 4 \Phi K K^{ab} \nabla_a \nabla_b \Phi
+ (K^2 - 4 
{\cal R} ) \Phi \Delta  \Phi 
\nonumber \\
&+&  ( 2 {\cal R}^2 
+ 2 K^4 - 5 {\cal R} K^2 - 4 {\cal G}_{ab} K^{ab} K)
\Phi^2 ] + 2K \nabla_a {\cal V}^a_{(2)} [K] \,,
\end{eqnarray}
where ${\cal V}^a_{(2)} [K]$ is defined by (\ref{eq:va2}).
We isolate a total divergence in the final term 
using
\begin{equation}
2K \nabla_a {\cal V}^a_{(2)} [K]
= - 2 (\nabla_a K) {\cal V}^a_{(2)} [K] 
+  \nabla_a {\cal V}^a_{(2)} [K^2]  
\,,
\end{equation}
where we define
\begin{equation}
{\cal V}_{(2)}^a [K^2] =  2 \sqrt{g} \left[ \Delta\Phi \nabla^a \Phi -
\Phi \nabla^a \Delta
\Phi
+  \Phi (2 K^{ab} - K g^{ab}) \nabla_b \Phi \right]\,,
\end{equation}
to obtain
\begin{eqnarray}
\delta^2 \int dA \; K^2  
&=& \int dA \;  \Phi [ 
2  \Delta^2  
+ 4  K K^{ab} \nabla_a \nabla_b 
+ (K^2 - 4 
{\cal R} ) \Delta  +
 2 {\cal R}^2 
\nonumber \\
&-&
  2 (\nabla_a K) ( 2 K^{ab} - g^{ab} K )
\nabla_b  
+ 2 K^4 - 5 {\cal R} K^2  - 4 {\cal G}_{ab} K^{ab} K ] \Phi
 \,,
\end{eqnarray}
We can simplify this expression using
\begin{equation}
- 2 \int dA \Phi (\nabla_a K) ( 2 K^{ab} - g^{ab} K ) \nabla_b \Phi
= \int dA [
2 K^{ab} \nabla_a \nabla_b K  -  K \Delta K +
(\nabla^a K ) (\nabla_a K ) ] \Phi^2\,,
\end{equation}
to arrive at the final expression
\begin{eqnarray}
\delta^2 \int dA \; K^2  
&=& \int dA \;  \Phi \; [ 
2  \Delta^2  
+ 4  K K^{ab} \nabla_a \nabla_b 
+ (K^2 - 4 
{\cal R} ) \Delta   
+2 K^{ab} \nabla_a \nabla_b K  
\nonumber \\
&-& 
K \Delta K +
(\nabla^a K ) (\nabla_a K ) 
+ 2 {\cal R}^2 
+ 2 K^4 - 5 {\cal R} K^2  - 4 {\cal G}_{ab} K^{ab} K ] \Phi
 \,.
\label{eq:expression}
\end{eqnarray}

Therefore, for 
the second order variation of the 
Helfrich hamiltonian (\ref{eq:hami}),
as expressed in (\ref{eq:op}), we can write the local differential operator
${\cal L}$ as
\begin{equation}
{\cal L} = 2\alpha \Delta^2 + 2 A \Delta
+ 2 A^{ab} \nabla_a \nabla_b + 2 B\,,
\label{eq:oper}
\end{equation}
where 
\begin{eqnarray}
2 A &=& \alpha (  K^2 - 4 {\cal R} ) - 2 \beta K - \lambda\,,
\\
2 A^{ab} &=& 4 \alpha  K K^{ab} + 
2\beta K^{ab} \\
2B  &=& 
\alpha [ 2 {\cal R}^2 
+ 2 K^4 - 5 {\cal R} K^2  - 4 {\cal G}_{ab} K^{ab} K 
+2 K^{ab} \nabla_a \nabla_b K   \nonumber \\
&-&
K \Delta K +
(\nabla^a K ) (\nabla_a K ) ]
- 2\beta {\cal G}_{ab} K^{ab} + \lambda {\cal R} + P K \,. 
\end{eqnarray}
We note that $A^{ab}$ is symmetric. 
Furthermore, the operator ${\cal L}$ is always self-adjoint, by which we mean that
\begin{equation}
\int dA \,\Phi_1 {\cal L} \Phi_2 = \int dA \,\Phi_2 {\cal L} \Phi_1\,.
\end{equation}
Thus, the eigenvalues of ${\cal L}$ are assured to be real valued. 
To see this, consider the terms that could potentially spoil self-adjointness
originating in the contributions proportional to $\alpha$ with two derivatives of $\Phi$.
Let $\tilde A^{ab} = A^{ab} + g^{ab} A$. Now, 
\begin{eqnarray}
\Phi \tilde A^{ab} \nabla_a \nabla_b \Phi &=& \Phi\nabla_a (\tilde A^{ab} \nabla_b \Phi)
- \Phi\nabla_a \tilde A^{ab}\,  \nabla_b \Phi \nonumber \\
&=& \Phi\nabla_a (\tilde A^{ab}\, \nabla_b \Phi) + 
{1\over 2} \nabla_a \nabla_b \tilde A^{ab}\,\, \Phi^2 \,,
\end{eqnarray}
modulo a total divergence which we discard. In this form 
the operator $\tilde A^{ab} \nabla_a \nabla_b$ is manifestly self-adjoint: we
have 
\[
\Phi_1\nabla_a ( \tilde A^{ab}\, \nabla_b \,\Phi_2 ) 
= \Phi_2 \nabla_b ( \tilde A^{ab}\, \nabla_a\, \Phi_1) 
\]
modulo two total divergences. We can write
\begin{equation}
{\cal L} = 2\alpha \Delta^2 + 2 \nabla_a \tilde A^{ab} \nabla_b 
+ 2 B_1\,,
\label{eq:oper1}
\end{equation}
where 
\begin{equation}
2B_1  = 
\alpha [ 2 {\cal R}^2 
+ 2 K^4 - 5 {\cal R} K^2  - 4 {\cal G}_{ab} K^{ab} K 
+2 \Delta (K^2- {\cal R})]  
- 2\beta {\cal G}_{ab} K^{ab} + \lambda {\cal R} + P K \,. 
\end{equation}

We note that the expression Eq.(\ref{eq:expression})
agrees with the one obtained in
\cite{For:86,Kle:86}, in the terms containing derivatives of $\Phi$.
These previous studies were   focused on the
effect of short wavelength fluctuations, where only these terms 
contribute. 
We also note that the expression we have obtained disagrees 
with the one obtained  previously in \cite{Hel.OuY:89}. Agreement is
restored in the special case of spherical and cylindrical configurations.

It is clear that a rigid translation of the surface should not alter the
second variation, at equilibrium. This provides a non-trivial check 
of our expression (\ref{eq:oper}). As it is not entirely obvious,
we will outline the details. Throughout, we will use extensively the Gauss-Weingarten 
equations (\ref{eq:gw1}), (\ref{eq:gw2}), and the contracted
Gauss-Codazzi-Mainardi
equations (\ref{eq:gausscc}), (\ref{eq:gaussc}), (\ref{eq:cmc}).
 Consider a constant normal deformation $\delta {\bf X} = 
{\bf a}$, such that ${\bf a} \cdot {\bf e}_a = 0$. We have for the
derivatives of $\Phi$ that appear in (\ref{eq:oper}), 
\begin{eqnarray} 
\nabla_a \nabla_b \Phi &=& - K_{ac} K^c{}_b ({\bf n \cdot a})\,,
\nonumber \\
\Delta^2 \Phi &=& [- 2 K^{ab} \nabla_a \nabla_b \Phi - (\nabla_a K ) (\nabla^a K ) 
- \Delta ( K_{ab} K^{ab} ) + (K_{ab} K^{ab})^2 ]({\bf n \cdot a})\,.
\nonumber 
\end{eqnarray}
If we now insert these expressions in the second variation, we find 
\begin{eqnarray}
\delta^2 F &=& \int dA\; \{ \alpha [ - 2 K^{ab} \nabla_a \nabla_b \Phi - (\nabla_a K ) (\nabla^a K ) 
- \Delta ( K_{ab} K^{ab} ) - K \Delta K  + 2 {\cal R} K^2 - K^4  ]
\nonumber \\
&+& \beta {\cal R} K + \lambda K^2 + P K \}
( {\bf n \cdot a})^2\,.
\end{eqnarray}
We now integrate by parts the terms that involve derivatives of the extrinsic curvature.
The desired simplifications follow from the crucial identity, for any surface vector $V^a$,
\[
\int dA \; (\nabla_a V^a ) ({\bf n \cdot a})^2 = 0\,,
\]
up to boundary terms. This is a consequence of 
$\nabla_a ({\bf n \cdot a}) = K_{ab} ({\bf e}_a \cdot {\bf a}) = 0$, by hypothesis.
Using this identity, we easily arrive at
\begin{equation}
\delta^2 F = \int dA \; K {\cal E} ({\bf n \cdot a})^2\,,
\end{equation} 
where ${\cal E}$ is the Euler-lagrange derivative
defined by (\ref{eq:shape}). Therefore at equilibrium the second variation
vanishes for rigid translations.

Note that a similar check has been used for the second variation of
particular configurations, as, for example, spheres in \cite{Pet:85,Hel.OuY:87}.
However, not all of the terms appearing in the second variation can be checked
considering only symmetric configurations, since many terms simply vanish identically
in these limits.

\section{Concluding remarks}

In this paper we have presented a covariant geometric approach
for examining the variation of geometric models of lipid membranes. For
concreteness, we have restricted our attention to the first and
second variation of the rigid bilayer couple  hamiltonian (\ref{eq:hami}). 
We note that it is straightforward to specialize the second variation in
the form (\ref{eq:oper}) to axisymmetric configurations.
Moreover, our approach can be extended straightforwardly to other geometric models. 
However, computational difficulties are to be expected if the membrane
has a free edge and it is no longer legitimate to throw away boundary terms.
When considering higher order variations, a different difficulty arises.
It is no longer justified to neglect tangential deformations. We will
address this issue in a future publication.

\vspace{1cm}

\noindent {\bf Acknowledgements}

\vspace{.5cm}

RC is partially supported by CONACyT under grant 32187-E.
JG and JAS are partially supported by  DGAPA at UNAM under grant IN119799.


\end{document}